\documentclass[a4paper,11pt]{article}
\pdfoutput=1 

\usepackage{jcappub,aas_macros} 

\usepackage[T1]{fontenc} 
\usepackage{array}
\title{\bf The CTA Sensitivity to Lorentz-Violating Effects on the Gamma-Ray Horizon}

\author[a]{M. Fairbairn,}
\author[b]{A. Nilsson,}
\author[a,c]{J. Ellis,}
\author[d]{J. Hinton,}
\author[d]{R. White}

\affiliation[a]{Theoretical Particle Physics and Cosmology Group, Physics Department, \\
King's College London, Strand, London WC2R 2LS, UK}
\affiliation[b]{Engineering Physics, Lund University, S\"{o}lvegatan 27, Lund, Sweden}
\affiliation[c]{Theory Division, Physics Department, CERN, CH-1211, Geneva 23, Switzerland}
\affiliation[d]{University of Leicester, University Road, Leicester LE1 7RH, UK}

\emailAdd{malcolm.fairbairn@kcl.ac.uk, \\ ~~~~~~~~~~~atf10ani@student.lu.se, \\ ~~~~~~~~~~~John.Ellis@cern.ch, \\ 
~~~~~~~~~~~jah85@leicester.ac.uk, \\ ~~~~~~~~~~~rw141@leicester.ac.uk}

\abstract{The arrival of TeV-energy photons from distant galaxies is 
expected to be affected by their QED interaction with intergalactic radiation fields 
through electron-positron pair production. In theories where high-energy photons violate Lorentz symmetry, 
the kinematics of the process $\gamma + \gamma\rightarrow e^+ + e^-$ is altered and the cross-section suppressed.
Consequently, one would expect more of the highest-energy photons to arrive if QED is modified by Lorentz violation than if it is not. 
We estimate the sensitivity of Cherenkov Telescope Array (CTA) to changes in the $\gamma$-ray horizon 
of the Universe due to Lorentz violation, and find that it should be competitive with other leading constraints.\\
~\\
KCL-PH-TH/2014-05, LCTS/2014-04, CERN-PH-TH/2014-017}

\begin{document}

 \maketitle
\section{Introduction}

It is widely expected that, at the quantum level, space-time should be regarded as a non-trivial medium,
the hypothesis named as `space-time foam' by J.A. Wheeler~\cite{Wheeler}. Motivated by this hypothesis,
it has been suggested in some approaches to quantum gravity that Lorentz invariance might break down at
high energies through effects $\propto (E/M_{LVn})^n$, where $n = 1, 2$ are the cases most often 
considered~\cite{amelino_1997,amelino_1998,colladay}.
For example, in~\cite{amelino_1997,amelino_1998,ellis_99} it was suggested that quantum gravity might
lead to a modified dispersion relation for $\gamma$-rays, with higher-energy photons propagating more
slowly than their lower-energy counterparts. This would lead to an energy-dependent refractive index
for light {\it in vacuo}, reminiscent of the effect on light of interactions with quantum modes of excitation
in a conventional medium. A related possibility, suggested in~\cite{Gambini:1998it}, is
that photons of different helicities might propagate at different velocities (cosmic birefringence). It has also been
suggested that the propagation of $\gamma$-rays  with the same energy might exhibit stochastic fluctuations in
their velocities~\cite{Ellis:1999uh}. These possible forms of Lorentz violation can be probed by measurements
of the arrival times of $\gamma$-rays from distant astrophysical sources that exhibit time structures in their
emissions, which provide some of the most sensitive probes of special relativity and Lorentz invariance. 
The sensitivities of current probes of the velocities of $\gamma$-rays from gamma-ray bursts (GRBs)
are $M_{LV1}=9.23 \times 10^{19}$~GeV and $M_{LV2}=1.3\times 10^{11}$~GeV (from Fermi \cite{fermilv}), whereas probes using active galactic nuclei (AGNs)
have sensitivities $M_{LV1} = 2.1\times 10^{18}$~GeV and $M_{LV2}=6.2\times 10^{10}$~GeV \cite{HESS:2011aa}.

In view of the importance of any possible future claim to have observed Lorentz violation, it is important to develop alternative probes
of possible effects, e.g., in order to avoid systematic errors such as issues that arise when one
tries to disentangle effects on $\gamma$-ray propagation from time delays and dispersions imprinted at the sources.

An attractive possibility is offered by the fact that the modified dispersion relation for $\gamma$-rays
that could be induced by Lorentz violation would affect
the kinematics for the production of $e^+ e^-$ pairs in photon-photon collisions, leading to an \emph{increase} in optical depth for
very-high-energy (VHE) $\gamma$-rays, an effect which would be more pronounced for higher-energy 
photons~\cite{kifune,protheroe,ellis_2000,Stecker:2001vb,Stecker:2003pw}. Put simply, VHE $\gamma$-rays that would normally {\it not} 
arrive from distant AGNs or GRBs could become detectable because Lorentz violation would suppress
the pair-production of electrons and positrons that would otherwise have stopped them from getting through.  (For a recent parallel study of the effect of Lorentz violation on the interaction with VHE gamma rays in the atmosphere, see \cite{Rubtsov:2013wwa}.)

VHE $\gamma$-rays traveling cosmological distances encounter low-energy photons belonging to the 
cosmic microwave background (CMB) and the extragalactic background light (EBL) due to starlight, 
both primary and re-emitted after absorption by dust~\cite{magicopac,hessopac,fermiopac}. At sufficiently high energies, 
according to conventional quantum electrodynamics (QED) some fraction of the emitted $\gamma$-rays are absorbed 
as they scatter off the EBL through pair production of an electron and a positron: 
$\gamma_{VHE} + \gamma_{EBL} \rightarrow e^+ + e^-$~\cite{gould2}. This endows intergalactic space with an effective `opacity', 
and the mean free path for a photon of given energy traveling cosmological distances can be calculated in QED,
leading to an energy-dependent horizon for the possible origin of VHE $\gamma$-rays that reach our Galaxy \cite{gould1}.
At even higher energies, interactions with the CMB become possible, at which point this horizon becomes much closer,
because of the very large number density of these microwave photons relative to the EBL (see, e.g., Fig.~5. in~\cite{fairbairntroitsky}).

On the other hand, if VHE $\gamma$-rays have a Lorentz-violating dispersion relation, the processes
$\gamma_{VHE} + \gamma_{EBL, CMB} \rightarrow e^+ + e^-$ may become kinematically forbidden,
in which case some VHE $\gamma$-rays could reach our Galaxy from beyond the conventional horizon~\footnote{There
may also be effects on the absorption cross-section due to (stochastic) energy non-conservation during the
$\gamma \gamma \to e^+ e^-$ interaction itself~\cite{ellis_2000},  
which could (partially) nullify the effects of the modified $\gamma$-ray
dispersion relation, but we do not discuss such a possibility here.}.
This article discusses the possibility of detecting these $\gamma$-rays with the soon-to-be built 
Cherenkov Telescope Array (CTA), a possibility discussed previously in~\cite{Ellis:2011ek}. 
The new array will have greater sensitivity and effective area than
previous experiments, with an array of widely-spaced small-sized telescopes (SSTs) to extend the sensitivity to the highest energies, up to several hundred TeV~\cite{ctamonte}. This makes it an excellent laboratory to search for the 
effects of Lorentz violation on $e^+ e^-$ pair-production. The energies of the photons of interest to us
are around a hundred TeV, extending up to 200 TeV, such as may be emitted by AGNs~\footnote{Similar effects would apply to $\gamma$-rays from GRBs. These have been observed out to larger redshifts, but typically have lower energies that are less interesting for studying these possible effects of Lorentz violation.}. 

It should be noted that if CTA does detect such photons, they will be some of the highest energy 
$\gamma$-rays that will have ever been detected from an astronomical object or indeed observed in 
any kind of scientific experiment. It is thought that there are VHE photons
in the high-energy cosmic-ray spectrum, but to date there are only constraints on such photons~\cite{UHECRgamma}
and one recent claim of detection~\cite{sergey2013}. The results in this paper therefore probe physics at the current frontier of photon energy - more than an order of magnitude greater than the LHC. The huge opacity of the CMB for such photons
suggests that this energy range may mark the absolute energy limit of primary photons that
can arrive successfully from cosmological sources.  Conversely, if Lorentz violation
does show up, we may be able to observe much higher-energy photons from cosmological sources, 
assuming the huge astrophysical accelerators in AGNs are able to produce them.

The structure of this paper is as follows. In Section~\ref{secrad} we describe how we calculate the radiation fields
through which we the $\gamma$-rays must propagate, before presenting the cross-section for the interaction between
high-energy and background photons in the presence of Lorentz violation in Section~\ref{cs}.
Then, in Section~\ref{method} we outline our method for obtaining the sensitivity of a possible CTA
constraint, before listing the properties of the spectra we are testing and listing the constraints.
Finally, we present and discuss our results in Sections~\ref{results} and \ref{conc}.

\section{Mean Free Path of Lorentz-Violating High-Energy Photons}

In this Section we first describe our estimate of the spectrum of photons through which the $\gamma$-rays of interest
must propagate, and then the cross-section for electron-positron pair production in the presence of Lorentz violation,
before combining these elements with the source spectra in the following Section to explore the possible constraints on
$M_{LV1}$ and $M_{LV2}$.

\subsection{Radiation fields}\label{secrad}

The EBL is made up of two components, light coming directly from stars,
which is in the optical and ultraviolet part of the spectrum (0.1-1 eV), and secondary infra-red radiation associated 
with the same starlight, which has been absorbed on its way out of galaxies and re-emitted at lower energies ($10^{-3}-10^{-2}$ eV). 
While more energetic, the number density of these EBL photons is much lower than the number density of photons from the CMB, which have energies of around $10^{-4}$ eV.

We obtain the background light from stars using the method of Finke, Razzaque and Derner,
where the stellar background is modeled as a collection of black-body spectra integrated over 
time and mass function~\cite{finke} (see also, e.g., \cite{primack} for more elaborate and 
complete ways of calculating the EBL). This gives rise to a redshift-dependent radiation field
that depends upon the history of the star formation rate, which we fit to the data in~\cite{hopkins}. 
We add to the EBL the CMB, which is simply a perfect black-body spectrum.
Since the sources that we consider are relatively close on a cosmological scale, with $z<0.1$,
we do not need to consider the redshift evolution of the EBL and the $z=0$ flux is sufficient to  estimate the overall effect.
However, our code automatically includes redshift evolution of the EBL and CMB. 

\begin{figure}[!htbp]
\centering
\includegraphics[scale=0.5]{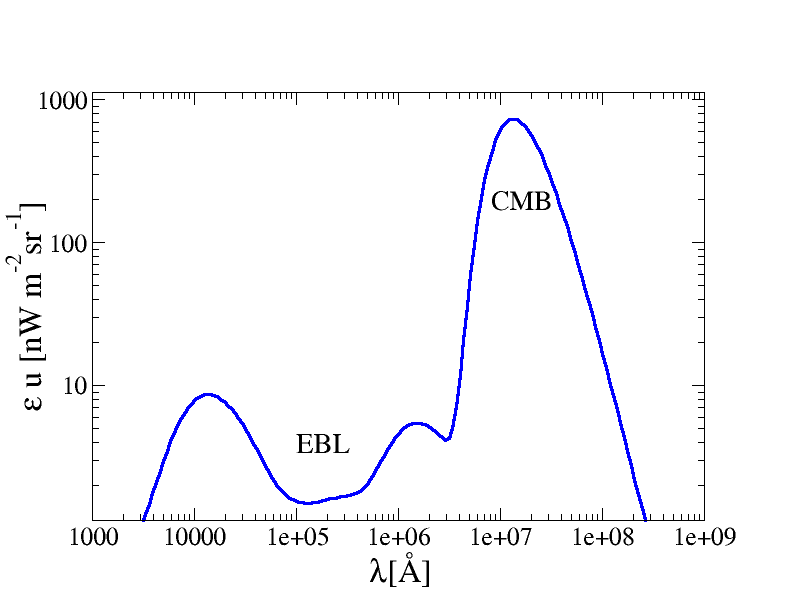}
\caption{\it Spectrum of the extragalactic background light (EBL) and the cosmic microwave background (CMB)
obtained as described in the text.}
\label{spec}
\end{figure}

The resulting spectrum is shown in Fig.~\ref{spec}. The precise form of the EBL spectrum, e.g.,
the position, shape and amplitude of each peak, is not critical for this analysis as the VHE ($E\sim$ 100 TeV)
$\gamma$-rays that are important for the analysis in this paper scatter mainly off the CMB, 
which is measured to extremely high precision~\cite{firas}.

\subsection{Pair-Production Cross Section with a Modified Dispersion Relation}\label{cs}

In this Section we outline the way that the cross section changes in the presence of Lorentz violation, 
mainly following the notation of Protheroe and Meyer~\cite{protheroe}. The parameter that controls the 
degree of Lorentz violation we write as $M_{LVn}: n = 1, 2, ...$ such that in the limit $M_{LVn} \to \infty$ one obtains
conventional special relativity, and one sees greater deviation from Lorentz symmetry as $M_{LVn}$ is reduced.

The modified photon dispersion relation when $M_{LVn}\ne\infty$ can be expressed kinematically in terms of
an effective mass of the photon~\cite{kifune} 
\begin{equation}
\beta_\gamma^2=1-\left(\frac{E_\gamma}{M_{LVn}}\right)^n \qquad ; \qquad m_\gamma^2= \frac{E_\gamma^{2+n}}{M_{LVn}^n},
\label{power}
\end{equation}
which vanishes in the limit of unbroken Lorentz symmetry. As mentioned in the Introduction, $n=1,2$ 
are the primary cases of interest, so these are the cases we study here. The cross-section for electron-positron pair production 
when $m_\gamma\ne0$ changes, since the normal integrals over photon energy have a different kinematically-allowed range: 
\begin{equation}
x_{\gamma \gamma}(E_\gamma)^{-1} = {1 \over 8 {E_\gamma}^2\beta_\gamma}
\int_{\varepsilon_{\rm min}}^{\infty} \, d\varepsilon
\frac{n(\varepsilon)} {\varepsilon^2} \int_{s_{\rm min}}^{s_{\rm
max}(\varepsilon,E_\gamma)} ds \, (s - m_\gamma^2(E_\gamma)c^4) \sigma(s) \, ,
\label{eq:mpl}
\end{equation}
where $n(\varepsilon)$ is the differential photon number density,
and $\sigma(s)$ is the total cross-section, with the centre-of-mass energy-squared being given by 
$s=m_\gamma^2(E_\gamma) + 2 \varepsilon E_\gamma(1 - \beta_\gamma\cos \theta)$,
where $\theta$ is the angle between the energetic photon ($\gamma$-ray) and the soft photon, 
$s_{\rm min} = (2 m_e)^2$, $\varepsilon_{\rm min} = (s_{\rm min}-m_\gamma^2(E_\gamma)) / [2E_\gamma(1+\beta_\gamma)]$, 
and $s_{\rm max}(\varepsilon,E_\gamma) = m_\gamma^2(E_\gamma) + 2 \varepsilon E_\gamma(1 + \beta_\gamma)$,
with $m_\gamma$ given by (\ref{power}).
In order to consider $e^+ e^-$ pair-production by $\gamma$-rays without Lorentz violation, 
we take $M_{LVn}=\infty$, so that $m_\gamma=0$ and $\beta_\gamma=1$, and we retrieve the normal
QED pair-production cross-section.

It is appropriate to mention at this point that the kinematics outlined above take into account the possibility of Lorentz violation for photons but not for electrons, which would also change the threshold dynamics for the pair production process.  The constraint on the Lorentz violating propagation of electrons is a lot more severe due to observations of high energy synchrotron radiation from the Crab Nebula which would have a distorted spectrum if high energy electrons had a modified dispersion relation \cite{Jacobson:2002ye}.

\subsection{Probability of Arrival for Different $M_{LVn}$}

\begin{figure}[!htbp]
\centering
\includegraphics[scale=0.5]{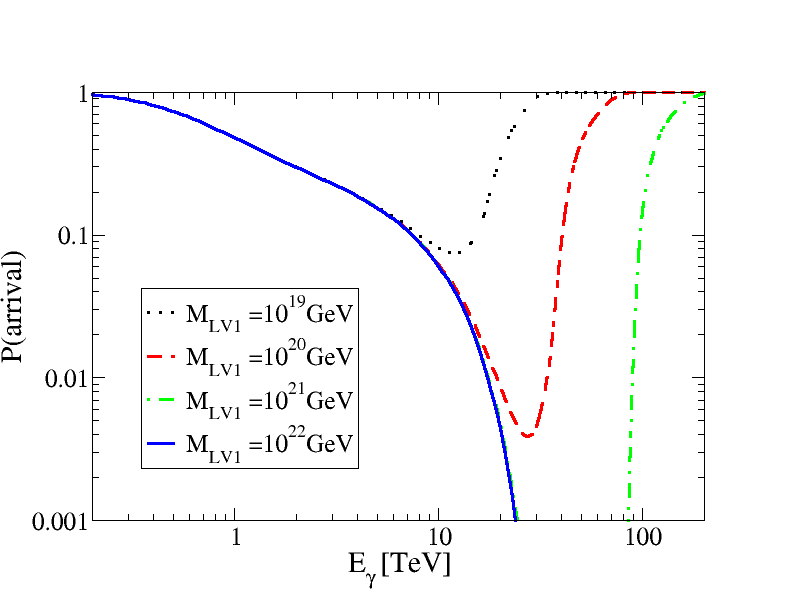}
\caption{\it The arrival probability of a photon emitted from a hypothetical source at redshift $z=0.05$ as a function of energy. 
The different curves represent different values of the Lorentz-violating scale $M_{LV1}$. 
VHE photons with energies $\gtrsim 100$~TeV can travel through the CMB effectively unimpeded.}
\label{prob}
\end{figure}

We are now in a position to combine the modified cross-section with the radiation fields described in
Section~\ref{secrad} to see the effect on the opacity of the Universe for energetic $\gamma$-rays
of lowering $M_{LVn}$ from infinity. Typical results are seen in Fig.~\ref{prob}. The solid blue line shows the survival probability for a $\gamma$-ray of the indicated energy to reach a detector after emission
from a source at redshift $z = 0.05$ in the case that $M_{LV1} = 10^{22}$~GeV, which is close to the limit $M_{LV1}\to\infty$
corresponding to the absence of Lorentz violation. 
We see that the arrival probability falls monotonically with energy, and is below 10\% for $E_\gamma \sim 10$~TeV.

The largest effect on the arrival probability shown in Fig.~\ref{prob} is for the lowest value of $M_{LV1}$ shown,
namely $M_{LV1} = 10^{19}$~GeV, indicated by the dotted black line. 
In this case, after dropping briefly below 10\% when $E_\gamma \sim 10$~TeV, 
the probability rises again for larger $E_\gamma$, becoming indistinguishable from unity when $E_\gamma \sim 40$~TeV.
The cases $M_{LV1} = 10^{20}, 10^{21}$~GeV (indicated by the dashed red and dot-dashed green lines, respectively)
are intermediate, with arrival probabilities rising towards unity at energies $E_\gamma \gtrsim 100$~TeV.

Fig.~\ref{prob} illustrates the principle of the analysis performed in this paper: in the presence of Lorentz
 violation, depending on the value of $M_{LVn}$, VHE $\gamma$-rays may be observable even if lower-energy
 $\gamma$-rays are not. This is potentially a clear signature of the form of Lorentz violation explored here.

\section{Choice of Astrophysical Sources and the Method used to Derive Constraints}\label{method}

Having established that the absence of dimming of VHE photons from astrophysical sources is a
potentially interesting signature of Lorentz violation, we now study the potential sensitivity of CTA to this
signature in the $E_\gamma$ range of interest, i.e., up to and around 100~TeV. In order to do this, we base
our analysis on the CTA spectrum sensitivity predictions in~\cite{ctamonte}. We assume that the spectrum of
a given AGN is obtained at low energies and then assume either (a) that it extrapolates all the way up to 200~TeV,
well into the range that cannot be observed due to the opacity of the EBL, or (b)
assume that there is some cut-off coming in at 10 or 20~TeV where the physics of the astrophysical
accelerator is no longer able to create as many photons. We will look at the constraints on Lorentz
violation that can be obtained under both assumptions.

\subsection{CTA Sensitivity and Statistical Method used}

We base our analysis on the tabulation of the expected values of the background from cosmic rays for 50 hours of observation 
within the point-spread function of CTA at that energy presented in Fig.~16 of~\cite{ctamonte} (observations at 20 degree zenith angle with array layout `I'). 
We then calculate the expected number of signal events
during the same period from the source, which varies at high energy, 
depending upon the energy scale of Lorentz violation $M_{LVn}$~\footnote{Another criterion for detection
would be to test whether the extrapolated spectrum crosses the 50 hours detection line given in
Fig.~8 of~\cite{ctamonte}.}. 

We consider only those high-energy bins in which the expected flux is less than the expected background
in the absence of Lorentz violation, before turning on the Lorentz violation by lowering the scale $M_{LVn}$ from infinity.
As we do so, the expected flux in those bins increases, and we calculate the probability of obtaining that flux in those bins
from background alone, assuming a Poisson distribution. We then calculate the product of the probabilities
in the different bins, using the appropriate penalty for combining multiple bins.
We look for the value of $M_{LVn}$ that gives a probability equivalent to a 5-$\sigma$ discovery, i.e., 
formally one chance in around 3.5 million that the expected signal could be a background fluctuation.
In the energy range of interest here the sensitivity curves~\cite{ctamonte} are limited 
by the requirement of a minimum of 10 signal events per bin. This is a reasonable indication of the flux limit 
for calculating a spectrum with 5 independent bins per decade but, as the background level in this energy range is very low, 
a significant detection (over several bins) is possible for much lower fluxes.

The limit given corresponds roughly to the 
detection of 9 signal events against a background of 1.6 in the energy range 25.1~TeV  to 251~TeV when we can use five bins,
or 11 events against a background of 2.4 in the energy range 15.8~TeV to 251~TeV when we can use six bins.
However, by calculating the Poisson probability in each bin separately and then by combining them with the appropriate penalty,
we are slightly more sensitive than this.

This method has the consequence that, as one decreases the energy scale of the assumed high-energy 
cut-off in the emitted flux, the constraint on $M_{LVn}$  becomes gradually weaker. However, 
the number of bins in which one would see no flux in the case of no Lorentz violation also increases, 
so the background that needs to be overcome increases in jumps, and constraints obtained in this way will reflect this non-smooth behaviour as one changes the energy of the cut-off.  More complex statistical tests could be devised and employed in the future that do not possess this weakness.

\subsection{Astrophysical Sources}

We have considered three VHE $\gamma$-ray sources at relatively low redshifts, 
each of which has a well-measured high-energy spectrum below the expected cut-off due to the EBL.

\emph{Markarian 421 (Mrk~421)} is a blazar at a low redshift of $z = 0.03$ (around $125$ Mpc) that
is, together with Mrk~501, one of the first discovered and most studied VHE blazars~\cite{albert_2008}. 
It has been found to have extreme flux variability, occasionally doubling its output in just 15 min~\cite{aharonian_2002}.
Its spectrum during a low flux state can be described accurately by a power law with an exponential cutoff at 
$1.44$~TeV~\cite{albert_2008}:
\begin{equation}
\frac{dN_{421}}{dE} = 1.57 \cdot 10^{-9} \left(\frac{E}{0.2 \text{ TeV}}\right)^{-2.2} \cdot ~ \exp\left({-~\frac{E}{1.44~{\rm TeV}}}\right) \, \text{cm}^{-2} \,\text{s}^{-1} \,\text{TeV}^{-1} \, .
\label{421_diff}
\end{equation}
For the purposes of this paper, it turns out that the spectrum of Markarian 421 is observed to fall too steeply above 1~TeV 
for a useful flux to be expected at 100~TeV. Too few photons are expected in the region of the spectrum that would be 
affected by an interesting scale of Lorentz violation, so that the constraint obtained from this object would be much
weaker than the other objects we will consider - somewhere in the region of $M_{LV1}\ge 10^{17}$ GeV. 
We therefore do not include it among the results shown in Table~\ref{restab}.

\emph{Messier 87 (M87)} is a very large elliptical galaxy located in the Virgo cluster at redshift  $z = 0.004360$~\cite{mei_2007}. 
It is a radio galaxy with signs of high time variability in its spectrum~\cite{m87_magic}. 
The spectrum can be described by the power law~\cite{m87_magic}:
\begin{equation}
\frac{dN_{M87}}{dE} = 7.1 \cdot 10^{-12} \left(\frac{E}{0.3 \text{ TeV}}\right)^{-2.21} \, \text{cm}^{-2} \,\text{s}^{-1} \,\text{TeV}^{-1} \, 
\label{m87_diff}
\end{equation}
and is therefore very bright at high energies.

Unfortunately, M87 turns out to be unsuitable since its relative proximity in cosmological terms 
(a distance of only around 18 Mpc) means that the opacity for even VHE photons is not large.
One therefore expects a considerable number of photons to arrive in the highest energy bins, even with no Lorentz violation.
The difference when Lorentz violation suppresses the opacity is therefore very difficult to detect and, because of this, 
M87 also turns out not to be of use for our analysis.

\emph{Markarian 501 (Mrk 501)} is a blazar located at $z = 0.034$ that is one of the most studied AGNs due to its high flux and low redshift~\cite{mrk501_q}. It is the brightest source in the sky above 100 GeV and was therefore one of the first blazars found to have VHE emission. Mrk501 displays high spectral variability, as well as short flares and long outbursts~\cite{mrk501_q, quinn_1996, catanese_1997}. 
The spectrum at low energies $E_\gamma \sim 0.2 - 3$~TeV can be described accurately by a power law,
as shown in~\cite{mrk501_q}:
\begin{equation}
\frac{dN_{501}}{dE} = 5.78 \cdot 10^{-12}  \left(\frac{E}{1 \text{ TeV}}\right)^{-2.72} \, \text{cm}^{-2} \,\text{s}^{-1} \,\text{TeV}^{-1} \, .
\label{501_diff}
\end{equation}
The spectrum is such that one might expect photons in the highest-energy bins in the absence of opacity,
but the distance of the source is such that interaction with the CMB would stop them from arriving.  This is the ideal situation since if Lorentz violation acts to suppress the opacity, one should be able to detect photons in the highest-energy bins.

The final statement above assumes that the power law (\ref{501_diff}) continues unbroken to 200~TeV, 
but it is clearly not realistic to assume that the spectrum can remain the same up to arbitrarily high energies.
Accordingly,  we consider two possible scenarios for its modification at high energies: an exponential cut-off at either 10~TeV
or 20~TeV.

\section{Results}\label{results}

\begin{figure}[!htbp]
\centering
\includegraphics[scale=0.6]{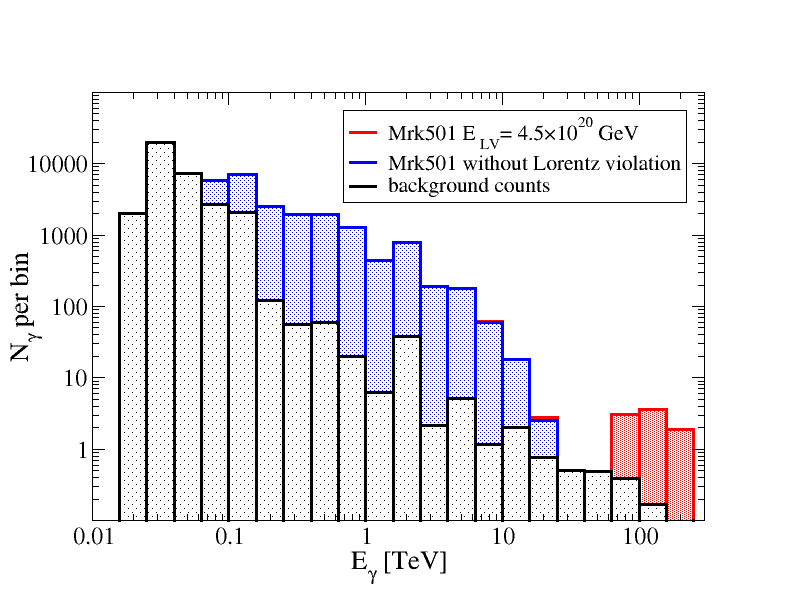}
\vspace{-0.7cm}
\caption{\it The expected number of signal events (blue and red columns) compared with the expected number
of background events alone (black columns), calculated for 50 hours of observation of the AGN Markarian~501,
assuming the power-law spectrum (\protect\ref{501_diff}).
The red columns represent the expected flux assuming a Lorentz-violating energy scale 
$M_{LV1}=4.5\times 10^{20}$~{\rm GeV}, whereas the blue columns denote the flux expected
in the absence of Lorentz violation, and are identical to the red columns below 15~TeV.}
\label{Mrk501}
\end{figure}

\begin{figure}[!h]
\centering
\includegraphics[scale=0.6]{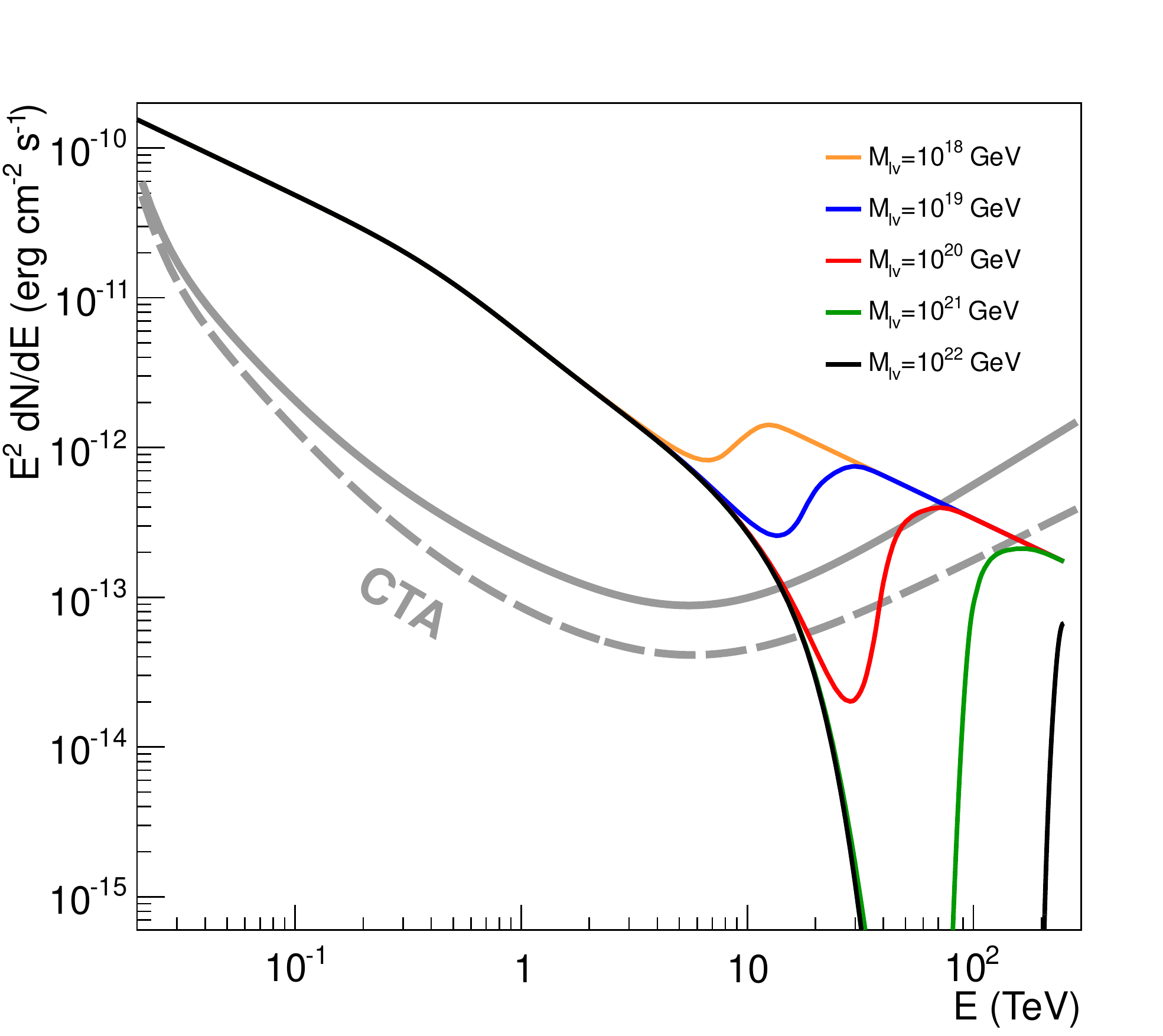}
\vspace{-0.3cm}
\caption{\it The expected spectrum of the AGN Markarian 501 for different values of $M_{LV1}$ 
vs the CTA sensitivity. The upper (solid) sensitivity curve is that presented in \cite{ctamonte},
whereas the lower (dashed) sensitivity curve uses wider energy bins and removes the requirement of 10 photons per bin as explained in the text.}
\label{dnde}
\end{figure}

\begin{table}
\vspace{1cm}
\centering
{\renewcommand{\arraystretch}{2}
\begin{tabular}{ | l | c | c | c | }
\hline
 & Power-law flux & 20-TeV cut-off & 10-TeV cut-off\\
 & $5\sigma$ $(3\sigma)$ [GeV]& $5\sigma$ $(3\sigma)$ [GeV]& $5\sigma$ $(3\sigma)$ [GeV]\\
\hline
  $n=1$ & $4.5\times 10^{20}$ $(1.4\times 10^{21})$& $1.8\times 10^{19}$ $(3.2\times 10^{19})$ & $4.1\times 10^{18}$ $(9.1\times 10^{18})$\\
\hline
  $n=2$ & $5.9\times 10^{12}$ $(1.2\times 10^{13})$& $5.1\times 10^{11}$ $(9.8\times 10^{11})$& $2.7\times 10^{11}$ $(4.2\times 10^{11})$\\
\hline
\end{tabular}}
\caption{\it The 5-$\sigma$(3-$\sigma$) sensitivities to $M_{LVn}$ for the cases
$n=1$ and $n=2$ in (\protect\ref{power}) estimated by combining the likelihoods in
high-energy bins as described in the text, assuming 50 hours of observations of the AGN
Markarian~501. In each row, the leftmost entry is obtained assuming a power-law
extrapolation of the emitted flux from low-energy data, and the centre and rightmost entries
assume exponential cut-offs at 20 and 10~TeV, respectively.}
\label{restab}
\end{table}

Fig.~\ref{Mrk501} shows the number of signal events from Markarian 501 expected after 50 hours of observation,
assuming a power-law spectrum
with no exponential cut-off. We plot the signal for the case of no Lorentz violation in blue, 
and with a Lorentz violation scale $M_{LV1}=4.5\times 10^{20}$~GeV in red.
Below 15~TeV the red and blue plots are indistinguishable, 
since Lorentz violation affects primarily the highest-energy photons,
and has negligible effects in these bins. We also plot for comparison the expected number of background events 
(black histograms), also calculated for 50 hours of observation. We see that the signal far exceeds the background at 
energies below $\sim 10$~TeV, but then dips below the background, before rising above
it again in the red Lorentz-violating case at energies $\gtrsim 100$~TeV. We combine the statistical
likelihoods that the background in the high-energy bins fluctuates as high as the red signal to calculate the 
5-$\sigma$ (3-$\sigma$) sensitivity to Lorentz violation~\footnote{At energies beyond those displayed, 
the estimated flux from Mrk~501 is too low to add significance to the observations.}, obtaining the results
shown in the upper row of Table~\ref{restab}. The lower row of Table~\ref{restab} shows the corresponding results
for $M_{LV2}$, the relevant parameter if Lorentz violation increases quadratically the energy.

Fig.~\ref{dnde} presents the same information in a different way, showing the differential spectra from 
the AGN Markarian 501 for different scales of $M_{LV1}$, 
again assuming a power law with no exponential high-energy cut-off. CTA differential sensitivity curves are presented for two scenarios. First the canonical 50~hr of observation and a 5-$\sigma$ detection criteria with 0.2 in log(E) bins and a minimum of 10 signal events required in each bin. As previously discussed this criteria is overly pessimistic for our purposes. Accordingly, a second sensitivity curve is added without the requirement for 10 signal events per bin, and wider bins of 0.6 in log(E). By observing where the spectrum crosses the sensitivity of CTA one can obtain a reasonable approximation to the more accurate numbers in Table~\ref{restab}.

The differential sensitivity line is explained earlier in the text, and it can be seen that even by observing
where the spectrum crosses the lines, one can obtain a reasonable approximation to the more accurate numbers in Table~\ref{restab}.

Table~\ref{restab} contains the fundamental result of this paper. It shows the values of $M_{LVn}$ 
required in order to obtain both 3-$\sigma$ and 5-$\sigma$ confidence level detections of VHE photons from 
Markarian~501 during 50 hours of exposure time for Lorentz-violating photon dispersion relations
with both $n=1$ and $n=2$. We consider three possibilities for both dispersion relations: 
that the spectrum continues as a power law to all the accessible energies, 
that the spectrum is cut-off exponentially at 20~TeV, and that the spectrum is cut off at 10~TeV.
In each case, we display the value of $M_{LVn}$ that would lead to a detection at the stated significance level.

\section{Conclusions}\label{conc}

We have shown in this paper that CTA potentially has very competitive sensitivity to
Lorentz violation, namely to $M_{LV1} \sim 10^{19}$~GeV and $M_{LV2}\sim10^{11}$ or more. The most promising
of the sources we have studied is Markarian~501. As shown in Table~\ref{restab}, assuming a power-law spectrum with no
exponential cut-off, we estimate a 5-$\sigma$ sensitivity to $M_{LV1} = 4.5\times 10^{20}$~GeV after 50 hours of observations,
assuming a continuing power-law emission flux.
Under the same assumptions, in the case of a quadratic modification of the photon
dispersion relation one could detect photons if $M_{LV2}=5.9\times 10^{12}$~GeV for $n=2$.

The projected constraints for both the linear $n=1$ and quadratic $n=2$ dispersion case are weaker than those already obtained from the
consideration of the non-observation of UHECR GZK photons in~\cite{sigl,Galaverni:2008yj}.  The two different constraints both have different model-dependencies, in the case of \cite{sigl} this is the unclear nature of the composition of cosmic rays at $10^{19}$ eV~\cite{composition}, a situation which may be clarified with Pierre Auger and Anita over the next years.  At the moment the proton content close to the GZK cut-off seems to be low, which casts doubt over the expected flux of UHECR GZK photons in Lorentz Vioating scnarios.  

In the case of the constraint obtained using the method presented in this paper, the model dependency is the uncertainty in the source spectra at very high energies.  It is therefore clear that understanding the emission spectra of the sources,
before the effect of any cosmic opacity, is critical in order to obtain truly reliable estimates of 
constraints on Lorentz violation. In order to do that, it would be desirable to have
better models of the engines responsible for the generation of VHE photons.
CTA will itself contribute to this understanding as it obtains much better statistics above the TeV scale.

The CTA performance curves used in this work (see Fig. \ref{dnde}) represent the baseline performance of CTA-South, 
which will have significantly better high-energy performance than that anticipated for the northern CTA installation. 
Mrk 501 is a northern source and would be observed from the southern site at zenith angles of $\sim60^{\circ}$, 
reducing the projected footprint of the area and hence collection area. However, the limits presented here 
(which require ~3 km$^2$ collection area at 100 TeV) can be reached if: a) the {\it goal} collection area of 
7~km$^2$ at $\sim20^{\circ}$ zenith angle can be reached for CTA-South, or b) CTA-North can be augmented with 
additional SST telescopes, or c) flares from a hard-spectrum southern hemisphere object reach the required fluxes. 
It seems plausible that one of these possibilities will transpire.

While CTA measurements of the energy spectra of the astrophysical sources considered here will
enable our estimates of the CTA sensitivities to Lorentz violation to be refined, our results
already indicate that observations of VHE $\gamma$-rays could provide an interesting
window on fundamental physics. We are unaware of any astrophysical source model that
could accommodate the reappearance of a source at very high energies following a loss of its
signal at lower energies caused by opacity due to the $\gamma \gamma \to e^+ e^-$ process.
This type of signal is arguably less ambiguous than studies of time delays in
the arrival times of $\gamma$-rays of different energies, and would be strong evidence for
new fundamental physics. On the other hand, we would not claim to be able to exclude the
possibility of some other fundamental physics mechanism that might diminish
the opacity of the Universe for VHE $\gamma$-rays (e.g., axion-like particles). It would therefore be desirable to
combine this type of study with other probes of modifications to our understanding of
fundamental physics. We would also like to emphasise that different modifications of
fundamental physics, such as energy non-conservation, might conspire to diminish or
even eliminate the effect discussed here.

Despite these caveats, we think that our analysis opens up a physics area of great potential interest for CTA, 
and look forward to the possibility of making more precise estimates of the sensitivity to new physics 
as our understanding both of the sources and of the performance of the final telescope itself improve.

\section*{Acknowledgments}
AN is extremely grateful for discussions with Konrad Bernl\"{o}hr.  The work of J.E. was supported in part by the London Centre for Terauniverse Studies (LCTS), using funding from the European Research Council via the Advanced Investigator Grant 267352. J.E., M.F., J.H. and R.W. are grateful for funding from the UK Science and Technology Facilities Council (STFC).

\bibliography{sources}{}

\end{document}